\documentclass{article}
\usepackage{amssymb}
\usepackage{amsbsy}

\usepackage{lineno}

\usepackage{comment}

\usepackage{lscape}

\usepackage{enumerate}
\usepackage{graphicx}
\newcommand{\be}{\begin{equation}}
\newcommand{\ee}{\end{equation}}

\def\bea{\begin{eqnarray}}
\def\eea{\end{eqnarray}}
\def\bean{\begin{eqnarray*}}
\def\eean{\end{eqnarray*}}

\newcommand{\barr}{\begin{array}}
\newcommand{\earr}{\end{array}}

\newcommand{\bed}{\begin{displaymath}}
\newcommand{\eed}{\end{displaymath}}
\newcommand{\bal}{\begin{array}{ll}}
\newcommand{\eal}{\end{array}}

\def\uplrarrow#1{\raise1.5ex\hbox{$\leftrightarrow$}\mkern-16.5mu #1}

\def\bvec#1{\raise1.5ex\hbox{$\rightarrow$}\mkern-16.5mu #1}

\def\m#1{\mathcal#1}

\begin{document}

\title{\hfill ~\\[-30mm]
       \hfill\mbox{\small }\\[30mm]
       \textbf{SuSy: The Early Years (1966-1976)}} 
\date{}
\author{\\        Pierre Ramond,\footnote{E-mail: {\tt ramond@phys.ufl.edu}}\\ \\
  \emph{\small{}Institute for Fundamental Theory, Department of Physics,}\\
  \emph{\small University of Florida, Gainesville, FL 32611, USA}}

\maketitle

\begin{abstract}
\noindent We describe the early evolution of theories with fermion-boson symmetry.
\end{abstract}
\thispagestyle{empty}
\vfill
\newpage
\setcounter{page}{1}

\section{Introduction}
\label{intro}
By the 1940's, physicists had identified two classes of ``elementary" particles with widely different group behavior, bosons and fermions. The prototypic boson is the photon which generates electromagnetic forces;  electrons, the essential constituents of matter, are fermions which satisfy Pauli's exclusion principle. This distinction was quickly extended to Yukawa's particle (boson), the generator of  Strong Interactions, and to nucleons (fermions). A compelling characterization followed: matter is built out of fermions, while forces are generated by bosons.  

Einstein's premature dream of unifying {\em all} constituents of the physical world should have provided a clue for that of fermions and bosons; yet it took physicists a long time to relate them by symmetry. This fermion-boson symmetry is called ``{\em Supersymmetry}".  

Supersymmetry, a necessary ingredient of string theory, turns out to have further remarkable formal properties when applied to local quantum field theory, by restricting its ultraviolet behavior, and  providing unexpected insights into its non-perturbative behavior. It  may also play a  pragmatic role as the glue that  explains the weakness of the elementary forces within the  Standard Model of Particle Physics at short distances.


\section{Early Hint}
In 1937, Eugene Wigner, with some help from his brother-in-law, publishes one of his many famous papers\cite{Wigner} ``On Unitary Representations of the Inhomogeneous Lorentz Group". He was then at the University of Wisconsin at Madison, a refugee from Princeton which had denied him tenure. It was not an easy paper to read, but its results were very simple: there were five types of representations labelled by the values of  $P^2\equiv p^\mu p_\mu=m^2$, one of the Poincar\'e group's Casimir operator. 
 
 All but two representations describe familiar particles found in Nature. Massive particles come with momentum $\bf p$, spin $j$, and $2j+1$ states of polarization, e.g.  electrons and nucleons with spin $1/2$. There are also four types of massless representations  with spin replaced by helicity (spin projection along the momentum). The first two describe massless particles with a single helicity (photons with helicity $\pm1$), or half-odd integer helicity, such as   ``massless" neutrinos with helicity $+1/2$.
 
The last two representations $O(\Xi)$ and $O'(\Xi)$ describe states which look like massless ``objects", particle-like in the sense that they have four-momentum, but with bizarre helicities: each representation contains an infinite tower of helicities, one with  integer helicities, the other with half-odd integer helicities. These have no analogues in Nature\footnote{``Infinite spin" representations do not appear in the Poincar\'e decomposition of the conformal group}.   
 
Physicists were slow in recognizing the importance of  group representations, even though Pauli's provided the first solution of the quantum-mechanical Hydrogen atom using  group-theory.  Wigner's paper does not seem to have moved any mountains, and infinite spin representations were simply ignored, except of course by Wigner. 

Yet, $O(\Xi)$ and $O'(\Xi)$ contained important information: they are ``supersymmetric partners'' of one another!

\section{Hadrons \& Mesons}
\label{sec:2}
Symmetries were gaining credence among physicists, not as a simplifying device but as a guide to the organization of Nature. Wigner and St\"uckelberg's ``supermultiplet model" unified $SU(2)$ isospin and spin. Once Gell-Mann and Ne'eman generalized isospin to $SU(3)$, it did not take long for 
 Feza G\"ursey and Luigi Radicati\cite{Gursey}, as well as Bunji Sakita\cite{Sakita}, to propose its unification with spin into $SU(6)$.  Pseudoscalar and vector mesons (bosons)  were found in the $\bf {35}$ of SU(6), while the hadrons (fermions) surprisingly lived in the $\bf{56}$, not in the $\bf{20}$\cite{Sakita}, as expected by the statistics of the time. This non-relativistic unification
 proved very successful, both experimentally and conceptually, since it led to the hitherto unsuspected {\em color} quantum number. 
 
In 1966, Hironari Miyazawa\cite{Miyazawa}  proposed further unification. His aim was to assemble the fermionic $\bf{56}$ and the bosonic $\bf{35}$ into one mathematical structure such as $SU(9)$, but at the cost of disregarding spin-statistics.

In order to explain the bounty of strange particle discovered in the 1950's, Sakata had proposed to explain mesons as $T\overline T$  bound states of the spin one-half triplet 
 
$$T~=~(\,p,\,n,\,\Lambda\,).$$ 
Miyazawa adds  a {\em pseudoscalar} triplet  

$$t~=~(\,K^+_{},  K^0_{},\,\eta\,),$$
to the Sakata spinor triplet. The hadron octet would then be described by another bound state, $T\bar t$, but he could not describe the spin three-half baryons decimet in the ${\bf 56}$. 
 
He introduces a toy model with two fundamental constituents, a spin one-half and a spin zero particle, $ {\bf p}=(\alpha_\uparrow,\alpha_\downarrow,\gamma)$. The nine currents
 
$$ {\bf p}^\dagger\lambda^{}_i{\bf p}=\cases{F^{}_i,\quad i=0,1,2,3,8;\cr G^{}_i,\quad i=4,5,6,7},$$ 
satisfy a current algebra with both commutators and anticommutators, 

\[
[\,F^{}_i\,,\,F^{}_j\,]~=~if^{}_{ijk} F^{}_k, \]
\[ [\,F^{}_i\,,\,G^{}_j\,]~=~if^{}_{ijk} G^{}_k,\]
\[ \{\,G^{}_i\,,\,G^{}_j\,\}~=~d^{}_{ijk} F^{}_k,\]
a ``generalized Jordan algebra" which he calls $V(3)$. This is the first example, albeit non-relativistic, of a superalgebra, today called $SU(2/1)$ with even part  $SU(2)\times U(1)$.

In 1967, he expanded his construction\cite{Miyazawa2}, to  general superalgebras he calls $V(n,m)$ with the idea of including the decimet. Alas, the phenomenology was not as compelling as that of $SU(6)$; two of the quarks inside a nucleon do not seem live together in an antitriplet color state.

In 1969, F.A. Berezin and G. I. Kac\cite{Berezin} show the mathematical consistency of graded Lie algebra which contains both commutators and anticommutators; they  give its simplest example generated by the three Pauli matrices $\sigma_+,\sigma_-,\sigma_3$. Physical applications are not discussed, although Berezin's advocacy of Grassmann variables in path integrals was no doubt a motivation.  

\section{Dual Resonance Models}
\label{sec:3}
In the 1960's, physicists had all but given up on a Lagrangian description of the Strong Interactions, to be replaced by the S-matrix program:  amplitudes were  determined from general principles and symmetries,  locality, causality, and Lorentz invariance.  Further requirements on the amplitudes such as Regge behavior and its consequent bootstrap program were still not sufficient to determine the amplitudes.  

In 1967, Dolen, Horn and Schmid\cite{Dolen} discovered a peculiar relation in $\pi-N$ scattering.  At tree-level, its fermionic $s$-channel  ($\pi\,N\rightarrow \pi\,N $) is dominated by resonances ($\Delta^{++}$, ...), as shown by countless experiments. On the other hand, its bosonic $t$-channel ($\pi\,\bar\pi \rightarrow N \,\overline N$) is dominated by the $\rho$-meson. Using the tools of S-matrix theory in the form of ``finite energy sum rules", they found that the Regge shadow of the bosonic $t$-channel's $\rho$-meson {\em averaged} the fermionic resonances in the $s$-channel! This was totally unexpected since these two contributions, described by different Feynman diagrams, should have been independent. Was this the additional piece of information needed to fully determine the amplitudes of Strong Interactions? This early example of fermion-boson kinship led, through an unlikely tortuous path, to modern Supersymmetry. 

An intense theoretical search for amplitudes where the $s$- and $t$-channel contributions are automatically related to one another followed. Under the spherical cow principle, spin was set aside and the search for DHS-type amplitudes focused on the purely bosonic process $\omega\rightarrow \pi\pi\pi$\cite{Ademollo}. Soon thereafter, Veneziano\cite{Veneziano} proposed a four-point amplitude with the desired crossing symmetry,

$$A(s,t) \sim \frac{\Gamma(-\alpha(s))\Gamma(-\alpha(t))}{\Gamma(-\alpha(s)-\alpha(t)},$$
where $\alpha(x)=\alpha_0+\alpha'x$  is the linear Regge trajectory. It displays an infinite number of poles in {\em both}  s-channel $s>0,~ t<0$ and t-channel $s<0,~t>0$. 
 
Veneziano's construction was quickly generalized to n-point ``dual" amplitudes. The infinite series of poles were recognized as the vibrations of a string\cite{string}. 

The amplitudes were linear combinations of tree chains which factorize into three-point vertices and propagators. A generalized coordinate emerged\cite{Fubini} from this analysis,

$$\quad Q^{}_\mu(\tau)=x^{}_\mu+\tau\,p^{}_\mu+\sum_{n=1}^\infty\frac{1}{\sqrt{2n\alpha'}}\left(a^{}_{n\mu}e^{in\tau}_{}-a^{\dagger }_{n\mu}e^{-in\tau}_{}\right),$$
with an infinite set of oscillators,

$$[a^{}_{n\mu},a^{\dagger}_{m\nu}]=\delta^{}_{nm}g^{}_{\mu\nu}$$
The vertex for emitting a particle of momentum $k_\mu$ from the linear chain was simple,

$$V(k,\tau)~=:e^{ik\cdot Q(\tau)}_{}:.$$
Out of its corresponding generalized momentum

\be P^{}_\mu(\tau)~=~\frac{dQ^{}_\mu}{d\tau}, \ee
one derived the operators,

$$L^{}_n~=~\frac{1}{2\pi}\int^\pi_{-\pi}d\tau e^{in\tau}_{}:P^\mu_{}P_\mu^{}:~\equiv~<:P^\mu_{}P_\mu^{}:>^{}_n, $$
which satisfy the Virasoro algebra\footnote{c-number is added anachronostically},
 
 $$[\,L^{}_m\,,\,L^{}_n\,]~=~(m-n)L^{}_{n+m}+{{\frac{D}{12}m(m^2-1)\delta^{}_{m,-n}}}.$$
 Its finite subalgebra, $L_0,L_\pm$, the Gliozzi algebra, generates conformal transformations in two dimensions. 
The  propagator was given by

$$\frac{1}{(\alpha'L^{}_0+1)}.$$


\section{Superstrings}
The Klein-Gordon equation for a point particle,

$$0~=~p^2_{}+m^2_{}~=~ <P^\mu_{}>^{}_0<P_\mu^{}>^{}_0+m^2, $$
could then be interpreted as a special case of 
 
$$0~=~ <P^\mu_{}P_\mu^{}>^{}_0+ m^2$$
suggesting a correspondence\cite{Ramond1} between point particles and dual amplitudes, 

$$<A><B>~\rightarrow~<A\,B>.
$$ 
Fermions should satisfy the Dirac equation, 

$$0~=~\gamma^{}_\mu\,p^\mu_{}+m~=~<\Gamma_\mu^{}>^{}_0<P^\mu>^{}_0+m.$$
This requires a generalization of the Dirac matrices  as dynamical operators,

$$\gamma^{}_\mu~~\rightarrow~~ \Gamma^{}_\mu~=~\gamma^{}_\mu+i\gamma^{}_5\sum_{n=0}^\infty\left(b^{}_{n\mu} e^{in\tau}_{}+b^{\dagger}_{n\mu} e^{-in\tau}_{}\right)
$$
where the oscillators are {\em Lorentz vectors}\footnote{Later was it realized that this made sense only in ten space-time dimensions where the little group is the spinor-vector schizophrenic $SO(8)$}, which satisfy anticommuting relations,

$$\{b^{}_{n\mu},b^{\dagger}_{n\mu}\}~=~\delta^{}_{nm}g^{}_{\mu\nu},$$
the sum running over the positive integers. 

This led me to propose the string Dirac equation in 
the winter of 1970\cite{Ramond}, which readily followed from that correspondence,

$$ 0~=~{{<\Gamma^{}_\mu\,P^\mu_{}>^{}_0+m}}.$$
The basic Dirac algebra, $\{\gamma\cdot p,\gamma\cdot p\}=p^2_{}$ 
is seen to be generalized to an algebra with both commutator and anticommutators,

$$\{\,F_n^{}\,,\,F^{}_m\,\}~=~ 2L^{}_{n+m},\quad [\,L^{}_n\,,\,F^{}_m\,]~=~(2m-n)F^{}_{m+n},$$
where $F_n=<\Gamma_\mu P^\mu>_n$, and these new $L_n$'s also satisfy the Virasoro algebra, but with a different $c$-number.

Andr\'e Neveu and John Schwarz then compute the amplitude for a dual fermion emitting three pseudoscalars with the Yukawa vertex, 
 
$$\Gamma_5^{}:e^{ik\cdot Q(\tau)}:,\quad \Gamma_5~=~\gamma_5(-1)^{\sum b_{n}^\dagger\cdot b_{n}},$$ 
and find that the resulting amplitude contains an infinite number of poles in its fermion-antifermion channel, and even identify the residue of the first pole\cite{Neveu2}!

 A new model with bosonic poles and vertices emerges, written in terms of  an infinite tower of anticommuting vector oscillators,

$$\{b^{}_{r\mu},b^\dagger_{s\nu}\}~=~\delta^{}_{rs}g^{}_{\mu\nu},\quad r,s={\textstyle\frac{1}{2},\frac{3}{2},\cdots}.$$
The triple boson vertex is given by

$$V_{NS}^{}(k,\tau)k_{}^\mu~=~ H^{}_\mu(\tau):e^{ik\cdot Q(\tau)}_{}:,$$
where

$$
H^{}_\mu(\tau)~=~\sum_{ r=1/2,3/2,\dots} [b^{}_{r\mu}e^{-ir\tau}+b^\dagger_{r\mu}e^{ir\tau}].$$
These are the building blocks of the  ``Dual Pion model"\cite{Neveu}, published in April 1971. The algebraic structure found in the generalized Dirac equation remains the same, producing a super-Virasoro algebra which decouples unwanted modes\cite{NST}, with $\Gamma_\mu$ replaced by $H_\mu$, through the operators,

$$G_r^{}~=~<H\cdot P>^{}_r,\quad r= {\textstyle\frac{1}{2},\frac{3}{2},\cdots}$$ 
 
The close relation of the two sectors is soon after formalized by Jean-Loup Gervais and Bunji Sakita\cite{Gervais} who write them in terms of  a world-sheet $\sigma$-model, with different boundary conditions, symmetric for the fermions, antisymmetric for the bosons.  They call the transformations  generated by the anticommuting Virasoro  opertors, {\em supergauge transformations}, the first time the name ``super" appears in this context.   

The following years saw the formulation of the RNS (NSR to some) ``Dual Fermion Model", generating  dual amplitudes  with bosons and fermions legs.
It lived in ten space-time dimensions, with states determined in terms of transverse fermionic and bosonic harmonic oscillator operators. 

In the fermionic ``R-sector", the spectrum of states is spanned by the fermionic ground state, $u|0>$ where $u$ is a fixed 32-dimensional spinor,  annihilated by both transverse  bosonic and fermionic oscillators, $a^{}_{ni}$ and $b^{}_{ni}$, $i=1,2\dots ,8$, and integer $n$. The fermion masses are determined by

$$\alpha'm^2_R~=~\sum_{n=1}^\infty n\Big[a^\dagger_n\cdot a^{}_n+b^\dagger_n\cdot b^{}_n\Big] $$
The  bosonic ``NS-sector"  spectrum starts with a tachyon,  $|0>$ annihilated by the same $a^{}_{ni}$, but also by the NS fermionic oscillators $b^{}_{ri}$, where $r$ runs over half-integers. The boson masses satisfy 

$$\alpha'm^2_{NS}~=~\sum_{n=1}^\infty n a^\dagger_n\cdot a^{}_n+\sum_{\scriptstyle r=\frac{1}{2}}rb^\dagger_r\cdot b^{}_r-\frac{1}{2}.$$

But there were idiosyncrasies. The correspondence between Neveu-Schwarz and the dual fermion states differed for states with an even number ($G\equiv(-1)^{\sum b_r^\dagger\cdot b^{}_r}=-1$) of $b^\dagger_r$, and states with an odd number, and there is a tachyon in the even number spectrum, at $\alpha'm^2_{NS}=-1/2$..  

In 1976, F. Gliozzi, Jo\"el Scherk, and David Olive\cite{GSO} noticed that the NS tachyon can be eliminated by requiring an odd number of anticommuting operators in the bosonic spectrum, ($G=-1$). The NS ground state 

$$\alpha'm^2_{NS}=0: ~~~b^\dagger_{1i}|0\rangle,$$
now consists of eight bosons, transforming as the vector(=spinor) $SO(8)$ representation.  The first excited states are 

$$\alpha'm^2_{NS}=1:~~~b^\dagger_{\scriptstyle\frac{1}{2}i}b^\dagger_{\scriptstyle\frac{1}{2}j}b^\dagger_{\scriptstyle\frac{1}{2}k}|0\rangle,~~b^\dagger_{\scriptstyle\frac{1}{2}i}a^\dagger_{1j}|0\rangle,~
b^\dagger_{\scriptstyle\frac{3}{2}i}|0\rangle,$$
that is $128=56(8.7.6/1.2.3)+64 (8.8)+8$ bosonic states, and so on.

In their next step, they  show that the R ground state solution could also be reduced to eight fermionic degrees of freedom. In ten dimensions, while a spinor has naturally thirty-two degrees of freedom, they showed that one can impose {\em both} chiral and Majorana (reality) restrictions on it, and reduce the spinor to eight dimensions, the spinor(=vector) $SO(8)$ representation. 

$$\alpha'm^2_{R}=0:~~~\psi_{\alpha}|0\rangle,~~\alpha=1,2\cdots 8.$$
The first excited state of the R-sector consists of

$$\alpha'm^2_{R}=1:~~~b^\dagger_{1i}\psi_{\alpha}|0\rangle,~~a^\dagger_{1j}\psi_{\alpha}|0\rangle,$$
with $128=8.8+8.8$ fermionic states! This was no accident, and using one of Jacobi's most obtuse relations, they showed that this equality obtained at all levels. Indeed this was supersymmetry, with the same number of bosons and fermions, albeit in ten space-time dimensions.   

Fermion-boson symmetry, born in its world-sheet realization, reappears as supersymmetry in ten-dimensional space-time. 

Meanwhile, behind the iron curtain, ...
\section{Russians}
\label{sec:4}
In March 1971, there appears a remarkable and terse paper by Yu. Gol'fand and E. Likhtman\cite{Golfand}  who extend the Poincar\'e algebra 
generated by $P_\mu$ and $M_{\mu\nu}$ to ``bispinor  generators", $W_\alpha$ and $\overline W_\beta$, which generate spinor translations.

Cognizant that spin-statistics  requires anticommutating spinors, they arrive at the parity-violating algebra,
  
\be
\{W,W\}~=~[P_\mu,P_\nu]~=~0,\quad \{W,\overline W\}~=~\frac{(1+\gamma_5)}{2}\gamma_\mu P_\mu.\ee
assuming no other subalgebra of the Poincar\'e group.  With  little stated motivation, they have written down the ${\cal N}=1$ superPoincar\'e algebra in four dimensions!
 
They identify its simplest representation: two ``scalar hermitean" fields $\phi(x)$ and $\omega(x)$, and one left-handed spinor field $\psi_1(x)$, of equal mass, the earliest mention of the Wess-Zumino supermultiplet. They do not consider auxiliary fields nor display the transformation properties of these fields. However, they show the spinor generators as bilinears in those fields, 

\be
W~=~\frac{(1+\gamma_5)}{2}\int d^3x\Big[\phi^*\uplrarrow\partial_0\psi^{}_1(x)+\omega(x)\uplrarrow\partial_0\psi^{c}_1(x)\Big].\ee

They also describe the  {\em massive} vector multiplet follows with the vector field $A_\mu(x)$, a scalar field $\chi(x)$ and a spinor field $\psi_2(x)$. They write down its  spinor current,

\be
W~=~\frac{(1+\gamma_5)}{2}\int d^3x\Big[\chi\uplrarrow\partial_0\psi^{}_2(x)+A_\mu(x)\uplrarrow\partial_0\gamma_\mu\psi^{}_2(x)\Big].\ee

This ground-breaking paper ends with the difficult task of writing interactions. Self-interactions of the WZ multiplet are not presented, 
 only  its interactions with a massive Abelian vector supermultiplet.  This,  the last formula in their paper, is a bit confusing since $\phi$ and $\omega$ now appear as complex fields (setting $\omega=0$ and replacing the complex $\phi$ by $\phi+i\omega$ is more what they need), but it contains now familiar features, such as the squared $D$-term. 

Gol'fand and Likhtman had firmly planted the flag of supersymmetry in four-dimensions. 

Interestingly, physicists on both sides of the iron curtain seemed  oblivious to this epochal paper. 

E. Likhtman seems to be the only one who followed up on this paper. He notices\cite{Lebedev} that the vacuum energy cancels out because of the equal number of  mass bosons and fermions with the same mass. He finds scalar masses  only logarithmically divergent, which he mentions in a later publication\cite{Likhtman}.   

In December 1972, in an equally impressive paper,  D.V. Volkov, and V.P. Akulov\cite{Volkov}, want to explain the masslessness of  neutrinos in terms of  an invariance principle. They note that the neutrino free Dirac equation is invariant under the transformations,

$$\psi\rightarrow \psi+ \zeta,\quad x_\mu\rightarrow x_\mu-\frac{a}{2i}(\zeta^\dagger\sigma_\mu\psi-\psi^\dagger\sigma_\mu\zeta),$$
where $\zeta$ is a global spinor. When added to the Poincar\'e generators, they form a group, of the type   Berezin and G. I. Kac's had advocated\cite{Berezin}  for algebras with commuting and anticommuting parameters. The translation of $\psi$ makes the neutrino akin to a Nambu-Goldstone particle with only derivative couplings. 

There follows a Lagrangian that describes its invariant interactions, which we can identify as a non-linear representation of supersymmetry.

The end of their paper contains this remarkable sentence ``We note that if one introduces gauge fields corresponding to the(se) transformations, then, as a consequence of the Higgs effect, a massive gauge field with spin $3/2$ arises, and the Goldstone particles with spin $1/2$ vanish". This remark is followed in October 1973, when  D. V. Volkov and V. A. Soroka\cite{Volkov2} generalize their transformations to local parameters and show explicitly that the fermionic Nambu-Goldstone particle indeed becomes a gauge artifact. Thus was born what became known as the ``Super Higgs Effect".

\section{Wess-Zumino}
In October 1973, Julius Wess and Bruno Zumino\cite{Wess}  generalize the world-sheet supergauge transformations of the RNS model to four dimensions. 

Theirs is the paper that launched the massive and systematic study of  supersymmetric field theories in four dimensions. 

The scalar (now called chiral or Wess-Zumino) multiplet is introduced.  It consists of two real scalar bosons, $A$ and $B$, a Weyl (Majorana) fermion $\psi$ and two auxiliary fields $F$ and $G$. Supergauge transformations generate the algebra,

\bea
\delta A&=&i\overline\alpha\psi,\quad \delta B=i\overline\alpha\gamma_5\psi,\nonumber\\
 \delta\psi&=&\partial_\mu(A-\gamma_5B)\gamma^\mu\alpha+n(A-\gamma_5B)\gamma_\mu\partial_\mu\alpha\nonumber\\
&&~+~F\alpha+G\gamma_5\alpha\nonumber\\
\delta F&=&i\overline\alpha\gamma^\mu\partial_\mu\psi+i(n-\frac{1}{2})\partial_\mu\overline\alpha\gamma^\mu\psi\nonumber\\
\delta G&=&i\overline\alpha\gamma_5\gamma^\mu\partial_\mu\psi+i(n-\frac{1}{2})\partial_\mu\overline\alpha\gamma_5\gamma^\mu\psi,\nonumber\\
&&\nonumber\eea
where $\alpha$ is an ``infinitesimal" anticommuting spinor, and $n$ is an integer assigned to the multiplet. With impressive algebraic strength, they are shown to close on both conformal and  chiral transformations. In particular, two transformations with parameters $\alpha_1$ and $\alpha_2$ result in a shift of $x_\mu$  by $i\overline\alpha_1\gamma_\mu\alpha_2$.

The free Lagrangian for the scalar multiplet follows,

$$
 {\m L}_{WZ}~=~-\frac{1}{2}\partial_\mu A\partial^\m A -\frac{1}{2}\partial_\mu B\partial^\mu B-\frac{i}{2}\overline\psi\gamma_\mu\partial^\mu\psi+\frac{1}{2}(F^2+G^2).
 $$
It is not invariant under supergauge transformations but since it transforms as a derivative, the action is invariant. In order to introduce invariant interactions, they derive the calculus necessary to produce covariant interactions, by assembling two scalar multiplets into a third, etc... .

They also introduce the vector supermultiplet, consisting of four scalar fields, $D$, $C$, $M$, $N$, a vector field $v_\mu$, and two spinor fields $\chi$ and $\lambda$, on which they  derive the supergauge transformations. By identifying the vector field with the chiral current generated by a scalar multiplet,

$$v_\mu^{}~=~B\partial_\mu A-A\partial_\mu B-\frac{1}{2}i\overline\psi\gamma_5\gamma_\mu\psi,$$
and following it through the algebra, they express all the vector multiplet fields  as quadratic combinations of the scalar supermultiplet. In particular $D=2{\m L}_{WZ}$. 

Finally, they notice that one can drop some of these fields, $C$, $N$, $M$, and $\chi$ without affecting the algebra (soon to be called the Wess-Zumino gauge), and write the vector multiplet Lagrangian in a very simple form,

$${\m L}_V~=~-\frac{1}{4}v^{}_{\mu\nu}v^{\mu\nu}_{}-\frac{1}{2}i\overline\lambda \gamma_\mu\partial^\mu\lambda+\frac{1}{2}D^2.$$
This paper contains many of the techniques that were soon to be used in deriving many of the magical properties of supersymmetric theories in four dimensions. 

In December 1973, Wess and Zumino present the one-loop analysis\cite{Wess2} of an interacting Wess-Zumino multiplet, and find remarkable regularities: the SuSy tree-level relations are not altered by quantum effects, the vertex correction is finite (leaving only finally where they find that only wave function renormalization), and finally that the quadratic divergences of the scalar and pseudoscalar fields cancel. 
As it was realized later, this addresses the ``gauge hierarchy problem", and strongly suggests SuSy's application to the Standard Model.
 \section{Representations }
The representations of the supersymmetry algebra were first systematically studied by Gell-Mann and Ne'eman (unpublished). They mapped the algebra in light-cone coordinates to one fermi oscillator, and found that in supersymmetry, the  massless representations of the Poincar\'e group assemble into two states with helicities separated by one-half, 

$$ (\lambda\pm \frac{1}{2}, \lambda),$$
and with the same light-like momentum, yielding an equal number of bosons and fermions.  The simplest is $\lambda=0$, with a real scalar and half a left-handed Weyl fermion. However, CPT-symmetric local field theories require the other half of the Weyl fermion, $({\textstyle \frac{1}{2}, 0)+(0, - \frac{1}{2}})$
which describe one Weyl fermion and a complex scalar boson, the ingredients of the Gol'fand-Likhtman-Wess-Zumino multiplet. 

The massless gauge supermultiplet,  ${\textstyle(1 , \frac{1}{2}) +(- \frac{1}{2},-1)},$
describes a gauge boson and its companion Weyl (Majorana) fermion, the gaugino.  

The supergravity supermultiplet, $ {\textstyle (2, \frac{3}{2})+(-\frac{3}{2},-2)}$
contains the graviton and the gravitino, remarkably the ingredients of interacting supergravity\cite{supergravity} 

They extend their analysis to the case of $\m N$ supersymmetries. Disregarding particles of spin higher than two, they find two cases with manifestly self-conjugate supermultiplets:  

$\m N=4$ supermultiplet, with helicities, 

$$
 {\textstyle(1)+4( \frac{1}{2})+6(0)+4(- \frac{1}{2})+(-1)},$$
and led in 1976 to the $\m N=4$ superYang-Mills theory\cite{N4}, with was found much later to have magical properties, such as an enhanced conformal symmetry, and ultraviolet finiteness! 

$\m N=8$ supergravity with helicities,

\bean
 {\textstyle(2)+8(\frac{3}{2})+28(1)+56( \frac{1}{2})+70(0)}+\\
+ {\textstyle56(- \frac{1}{2})+28(-1)+8( -\frac{3}{2})+(-2)},
\eean
which also led to a fully interacting theory, $\m N=8$ Supergravity\cite{N8}.
 
Massive representations of supersymmetry can be assembled using a group-theoretical Higgs mechanism. The massive vector representation contains a Dirac spinor, a massive vector, and a scalar particle,

$${\textstyle(1,\frac{1}{2})+(-1,-\frac{1}{2})+(0,-\frac{1}{2})+(0,\frac{1}{2})},$$
all of equal mass, as considered by Gol'fand and Likhtman.

\section{Towards the Supersymmetric Standard Model}
With the Wess-Zumino paper, the flood gates had been opened\cite{Ferrara}. In short order,  a supersymmetric version\cite{WZ3} of $QED$  is written down, with  Abelian gauge invariance, in which the Dirac electron spinor is accompanied by {\em two} complex spin zero fields. In January 1974, Abdus Salam and J. A. Strathdee\cite{Salam} assemble the fields within a supermultiplet into one superfield  with the help of anticommuting Grassmann variables. The same authors\cite{Salam2} coin the word ``super-symmetry" in a May 1974 paper which generalizes  supersymmetry to Non-Abelian gauge interactions. 

Before applying supersymmetry to the real world, several conceptual steps must be resolved. The absence of fermion-boson symmetry at low energies, requires it to  be broken. Secondly, its application to the electroweak theory demands the extension of  the Higgs mechanism. Finally the known particles must be assigned to supermultiplets.   
   
In 1974, Pierre Fayet and John Iliopoulos\cite{ILIO} produce the first paper on spontaneous breaking of supersymmetry in theories with a gauged Abelian symmetry by giving its $D$ auxiliary field a constant value. Their proposal is remarkably simple, just add to the Lagrangian for a $U(1)$ vector multiplet a $D$-term

$$
{\m L}^{FI}_V~=~{\m L}^{}_V~+~\xi D.$$
This extra term  violate neither Abelian gauge invariance, nor supergauge invariance, since its supergauge variation is a total derivative. The resulting field equation $<D>_0=\xi$ yields a theory where both gauge and supergauge invariances are broken.

A year later, Lochlainn O'Raifeartaigh\cite{O'R} invents a different way to spontaneous breaking of supersymmetry, in theories with 
several interacting scalar supermultiplets. Its simplest model involves three scalar supermultiplets, with equations of motion

$$
F_1^{}=-m\phi_2^*-2\lambda\phi_1^*\phi_3^*, ~ F_2=-m\phi_1^*,~ F_3=\lambda(M^2-{\phi_1^2}^*),
$$
where $m$, $M$ and $\lambda$ are parameters. There are no solutions for which all three $F_i$ vanish, and supersymmetry is broken. From these two early examples, the auxiliary fields are the order parameters of SuSy breaking.
 
Both schemes yielded an embarassing massless Goldstone spinor, which may have impeded the application of supersymmetry\footnote{In 1976, Weinberg and Gildener note that supersymmetry could explain a low mass scalar boson, but bemoan that it would produce a massless fermion!}. None of these authors  were aware of Volkov's papers. 

The second hurdle is the generalization of the Higgs mechanism to supersymmetry. This is done in the context of an unusual model by Pierre Fayet\cite{Fayet} in December 1974. Like Volkov and Akulov before, Fayet  builds models where the electron neutrino is the Goldstone spinor from the breakdown of supersymmetry\footnote{In 1974,  the Standard Model was not yet ``standard", and many authors were still presenting alternatives}, using the FI mechanism. 

Although the model building in this paper did not survive the test of time, two important and more permanent concepts  emerged. One is that the Higgs mechanism applies, but  {\em two} scalar supermultiplets are needed to achieve $SU(2)\times U(1)\rightarrow U(1)$ electroweak breaking, in accord with the number of surviving scalars in the massive vector supermultiplets. Also the existence of  $R$-symmetry, a new kind of continuous symmetry acting on both the fields and the Grassmann parameters of the superfields.  

It was not until July 1976, that Pierre Fayet\cite{Fayet2} generalizes the Weinberg-Salam (soon to be Weinberg-Salam-Glashow, and then Standard) model to SuSy. Its distinctive feature are:
\begin{itemize}

\item Two scalar superfields, $S$, $T$, (today's $H_{u,d}$) for EW breaking

\item Leptons and quarks are the fermions inside scalar supermultiplet. 

\item  A continuous $R$-symmetry

 \end{itemize}
The particle content is that the ``minimal supersymmetric model" (MSSM). Some kinks still need to be ironed out. having to do with  SuSy breaking ( {\em \` a la} Fayet-Iliopoulos in this paper), which produces a massless Goldstone spinor. The continuous $R$-symmetry in this paper behaves like a ``leptonic" number, but it prevents the spinor gluons from acquiring a mass. 

Today we know that SuSy breaking is an active area of theoretical research, even without the presence of a Goldstone fermion, eaten by the Super-Higgs mechanism.

\section{SuSy Today}
By stopping this history of fermion-boson symmetry in 1976, we rob the reader of the many wonderful concepts since discovered, but they are more than adequately covered in the 
 articles in this volume.

The seeds of today's Susy research were planted in these early papers. 

Almost forty years later, superstring theories have blossomed into a dazzling array of connected theories; the study of $\m N=4$ superYang-Mills theories is an active field of research, as is the possible finiteness of $\m N=8$ supergravity. 

The Hamiltonian is no longer fundamental, but derived from translations along SuSy's fermionic dimensions.


Few doubt of the existence of a deeper connection between bosons and fermions, but opinions differ at which scale it will be revealed: the breaking of Supersymmetry remains as mysterious as ever. 

Yet, the recent discovery of a low mass Higgs suggests that the universe displays more symmetry at shorter distances. 

Today, SuSy is unfulfilled, beloved by theorists, but so far shunned by experiments. 

In the words of the late Sergio Fubini, {\em ``We  do not know if supersymmetry is just a beautiful painting to put on the wall, or something more"}.


\end{document}